\documentclass[12pt,letterpaper]{article} 
\pdfoutput=1
\usepackage[includeheadfoot,
            marginratio={1:1,2:3}, 
            width=412pt, 
            height=688pt,]{geometry}

\usepackage{amsmath}
\usepackage{amsfonts}
\usepackage{amssymb}
\usepackage{stmaryrd}
\usepackage{graphicx}
\usepackage{cite}
\usepackage{hyperref}

\newcommand{\nc}{\newcommand}
\nc{\lb}{\llbracket}
\nc{\rb}{\rrbracket}
\nc{\gl}{\llbracket}
\nc{\gr}{\rrbracket}

\newcommand{\eq}[1]{\begin{equation}
                     \begin{split} #1 \end{split}
                     \end{equation}}

\newcommand{\ov}{\overline}

\newcommand{\op}{\hspace{1pt}}

\allowdisplaybreaks[2]
\numberwithin{equation}{section}

\begin{document}

\vspace*{-1.5cm}
\begin{flushright}
  {\small
  MPP-2016-103\\
  }
\end{flushright}

\vspace{1.5cm}
\begin{center}
 {\LARGE String Moduli Stabilization at  the Conifold }
\vspace{0.4cm}

\end{center}

\vspace{0.35cm}
\begin{center}
  Ralph Blumenhagen,  Daniela Herschmann,  Florian Wolf
\end{center}

\vspace{0.1cm}
\begin{center} 
\emph{ Max-Planck-Institut f\"ur Physik (Werner-Heisenberg-Institut), \\ 
   F\"ohringer Ring 6,  80805 M\"unchen, Germany } \\[0.1cm] 
\vspace{0.25cm}

\vspace{0.2cm}

 \vspace{0.5cm} 
\end{center} 

\vspace{1cm}


\begin{abstract}
We study moduli stabilization for type IIB orientifolds compactified on
Calabi-Yau threefolds in the region  close to conifold singularities
in the complex structure moduli space.
The form of the periods implies  new phenomena
like exponential mass hierarchies even in the regime of negligible
warping.
Integrating out the heavy conic complex structure modulus leads to an effective
flux induced potential for the axio-dilaton and the remaining complex
structure moduli containing
exponentially suppressed terms that imitate non-perturbative effects.
It is shown that this scenario can be naturally combined with the  large volume
scenario so that all moduli are  dynamically stabilized in the
dilute flux regime. As an application of this moduli stabilization
scheme, a string inspired model of aligned inflation is designed
that features a parametrically controlled   hierarchy of mass scales. 
\end{abstract}

\clearpage

\tableofcontents



\section{Introduction}
\label{sec:intro}

Moduli stabilization is one of the most important challenges  to
relate compactifications of string theory to our four-dimensional
world. Despite its importance, we think it is fair so say that
comparably few concrete scenarios have been discussed in the
literature. The ingredients used are tree-level fluxes as well as 
perturbative and non-perturbative corrections to the leading order
quantities. 

The work of Giddings-Kachru-Polchinski (GKP) \cite{Giddings:2001yu} revealed that Type IIB compactifications on
warped Calabi-Yau  threefolds equipped with localized sources and NS-NS and R-R
three-form fluxes are leading order solutions to the string equations
of motion. There, a scalar potential for the complex structure and the axio-dilaton moduli
is generated, while due to its no-scale structure, the K\"ahler moduli
remain as massless moduli. It is from this scenario that the landscape
idea of string vacua arose. In a second step, by turning
on non-perturbative corrections to the superpotential, the no-scale
structure was broken and the K\"ahler moduli could be stabilized.
This led to the KKLT \cite{Kachru:2003aw} and the large volume scenario (LVS) \cite{Balasubramanian:2005zx}. 

In \cite{Giddings:2001yu} it was pointed out that one can dynamically freeze the
complex structure moduli in the vicinity of a conifold singularity.
For that purpose, as for the Klebanov-Strassler throat \cite{Klebanov:2000hb}, one
turns on a three-form flux on the three-cycle that vanishes at the conifold
locus and an additional flux on its symplectic dual three-cycle.
In this case, at the tip of the throat the warp factor becomes large,
which leads to red-shifted masses of the modes localized there.
From the statistical analysis \cite{Ashok:2003gk} it even followed that the number
of vacua enhances close to a conifold locus. This was explicitly
verified for a concrete example in \cite{Conlon:2004ds}.

In this respect it is important to keep in mind that this analysis
was employing the usual effective supergravity action described
by the leading order K\"ahler potential and the Gukov-Vafa-Witten (GVW)
superpotential \cite{Gukov:1999ya,Taylor:1999ii}. 
In the case of strong warping, this action is not
any longer expected to be trustable as, due to red-shifting,
certain Kaluza-Klein modes can become light and one can have non-trivial
mixings among the modes. The effective action for warped
compactifications was studied in
\cite{DeWolfe:2002nn,deAlwis:2003sn,Giddings:2005ff,Shiu:2008ry}. 
If one wants to use the standard supergravity action one has to ensure that one
is working consistently in a dilute flux limit, where the backreaction
is suppressed.

In this paper we systematically study such flux compactifications close to a conifold
locus in the complex structure moduli space in the dilute limit.
Our main concern is to analyze the appearing moduli mass scales and
the  implications for string cosmology model building. 
It is a known and celebrated result  that, including a non-trivial warp factor, 
exponential hierarchies among the masses of the moduli are generated. 
This is true in particular for modes localized in a strongly warped throat.
As a matter of fact, this is 
the idea of the Randall-Sundrum scenario \cite{Randall:1999ee}. 
Inflationary models in warped throats with red-shifted inflaton masses
have recently been discussed in  \cite{Franco:2014hsa,Kooner:2015rza,Hebecker:2015tzo}.
In this paper we find that even in the
dilute flux limit, where the usual effective  supergravity theory is
applicable, exponential mass hierarchies between bulk modes are generated.

This observation is very interesting, as it allows to generate
exponential hierarchies in a controllable supergravity theory, a problem that 
was recently  of crucial relevance  for potential realizations of large-field inflation
in  concrete string theory set-ups. To build models with large
tensor-to-scalar ratios $r>0.01$ one needs a rolling of the inflaton
over trans-Planckian field ranges, which are hard to control in a
perturbation expansion (see e.g. \cite{Cicoli:2008gp} for a string
realization). 
Here the perturbative shift symmetry of axions 
does help and various models of axion inflation were
proposed (see e.g.
\cite{Freese:1990rb,Kim:2004rp,Dimopoulos:2005ac,Kaloper:2008fb,Silverstein:2008sg,McAllister:2008hb}
and \cite{Westphal:2014ana} for review). However, it turned out to  be quite a difficult task to
dynamically stabilize all the moduli at a higher mass scale than the inflaton
in a controllable way \cite{Blumenhagen:2014nba,Hebecker:2014kva,Blumenhagen:2015kja}. Moreover, such models of axion inflation came
under pressure also via the weak-gravity conjecture (WGC)\cite{Rudelius:2014wla,Rudelius:2015xta,Montero:2015ofa,Brown:2015iha}. 

Having this motivation in mind, in this paper we will proceed as  follows: In section 2 we briefly provide the main
ingredients for  Type IIB moduli stabilization, including  
geometric fluxes and instanton effects.
For the quintic we recall the form of the periods close to the
conifold singularity. 
The distinguished new issue is the appearance
of a logarithmic term for a certain period. Computing the
K\"ahler potential close to the conifold point, one realizes
the appearance of axion-like  shift-symmetries \cite{Garcia-Etxebarria:2014wla}, making
this regime interesting for realizing large field inflation. 
The latter has been subject of intense studies during the last two
years, where it has become clear that the weak-gravity conjecture and
concrete model building attempts  severely  constrain the
viability of such models.

In section 3, we first study 
the no-scale scalar potential for the conic complex structure modulus and
the axio-dilaton for the existence of minima close to the conifold
locus. These indeed exist in a controllable way  and for the mass of
the conic complex structure modulus we find the two salient features:
\begin{itemize}
\item{Independent of the overall size (K\"ahler modulus) of the Calabi-Yau, 
            it is exponentially larger than the mass of the axio-dilaton.}
\item{Requiring that its mass is lower than the string scale, leads to 
      a  size of the CY  that implies   the dilute flux
    limit, where warping can be neglected.}
\end{itemize}
The second point is very satisfying, as it means that the usual
supergravity action shows by itself its limitation to not strongly
warped configurations.
Integrating out the heavy conic  complex structure modulus,
one gets exponential terms in the axio-dilaton modulus that mimic
non-perturbative effects. Second, we study whether the above 
scenario can be combined with the large volume scenario to dynamically
provide an exponentially large overall volume of the CY.

Section 4 is devoted to possible cosmological applications of 
this conic LVS scenario. 
For having more complex structure moduli available, we 
generalize the evaluation of the periods to the mirror
of $\mathbb P_{11226}[12]^{(128,2)}$. 
Then the above mentioned exponentially
suppressed, instanton like contributions to the effective action
are investigated with respect to their potential to realize
a string inspired model of aligned axion inflation.
As  in
\cite{Blumenhagen:2014nba,Hebecker:2014kva,Kobayashi:2015aaa,Bizet:2016paj},
the inflaton is an axion-like complex structure modulus.
We also critically reflect the pre-assumptions made for constructing this model and discuss
the relation to a possible loop-hole in the mild version of the WGC.

\section{Type IIB fluxes  on CY threefolds}
\label{sec:framework}

As in many previous works, we consider the issue of moduli stabilization for type IIB string theory
compactified on orientifolds of Calabi-Yau threefolds $\mathcal M$ with
$O7$ and $O3$-planes.
Prior to turning on any fluxes, one obtains a moduli space 
parametrized by the vacuum expectation values of a set of scalar
fields invariant under the orientifold projection.

\subsection{Massless fields}

The orientifold projection splits the cohomology groups into
even and odd parts.
The resulting massless closed string moduli fields \cite{Grimm:2004uq}  of the effective four-dimen\-sio\-nal theory 
after compactification are summarized in table~\ref{table_moduli}.
\begin{table}[ht]
\centering
\renewcommand{\arraystretch}{1.3}
\begin{tabular}{|c|l@{\hspace{1pt}}l|c|}
  \hline
   number & \multicolumn{2}{c|}{modulus} &  name \\
  \hline\hline
  $1$ & $S$&$=e^{-\phi}-i\op C_0$ & axio-dilaton \\
  $h^{2,1}_-$ & $U^i$&$=v^i+i\op u^i$ & complex structure\\
 $h^{1,1}_+$ & $T_\alpha$&$=\tau_\alpha+ i \op \rho_\alpha$ & K\"ahler \\
 $h^{1,1}_-$ & $G^a$&$=S\op b^a+i\op c^a$ & axionic odd\\
\hline
     \end{tabular} 
     \caption{\small Moduli in type IIB orientifold compactifications.}
      \label{table_moduli}
\end{table}
The axionic odd moduli do not play any role in the subsequent
discussion so that we choose $h^{1,1}_-=0$. For convenience we
also choose $h^{2,1}_+=0$ so that we do not get any additional
abelian vector superfields.

The complex structure moduli $U^i$ are contained in the holomorphic three-form $\Omega_3$.
The latter can be expanded in the basis of three-forms 
as follows
\eq{
  \label{exp_02}
  \Omega_3 = X^{\lambda} \alpha_{\lambda} - F_{\lambda} \op\beta^{\lambda} \,.
}
where 
we denote a symplectic basis for the third cohomology of the Calabi-Yau manifold $\mathcal M$ by
\eq{
  \{\alpha_{\Lambda},\beta^{\Lambda}\} \in H^3(\mathcal M) \,, \hspace{60pt}
  \Lambda =0,\ldots, h^{2,1} \,,
}
and 
$X^{\lambda}$ and $F_{\lambda}$ are the periods of the Calabi-Yau
\eq{
                 X^{\lambda}=\int_{A^\lambda} \Omega_3\,,\qquad  
 F_{\lambda}=\int_{B_\lambda} \Omega_3\,.
}
Here  $A^\lambda,B_\lambda\in H_3({\cal M})$ denote  a basis of
Poincare dual three-cycles.
The $X^{\lambda}$ can be considered as homogeneous coordinates
of the complex structure moduli space. Inhomogeneous coordinates
are e.g. defined via $U^i=X^i/X^0$ with  $i = 1,\ldots,h^{2,1}$.

In special geometry, the periods $F_{\lambda}$ can be expressed as derivatives 
$F_{\lambda} = \partial F/\partial X^{\lambda}$ of a prepotential $F$. 
Recall that in the large complex structure limit the prepotential takes the form
\eq{
  \label{prepot}
   {F}=\frac{d_{ijk} \op{X}^i{X}^j{X}^k}{ {X}^0} \,, 
   \hspace{50pt}
   i = 1,\ldots,h^{2,1}\,,
} 
where the constants $d_{ijk}$ are the triple intersection numbers of
the  mirror dual CY.
At leading order in $\alpha'$, the K\"ahler potential for these chiral
superfields is given by
\eq{  K=-\log(S+\ov S) -2\log {\cal V} -\log \Big(-i \int \Omega_3\wedge
  \ov\Omega_3\Big)
}
where ${\cal V}$ is the total volume of the CY expressed in terms of
the four-cycle volumes ${\rm Re}(T_\alpha)$.

\subsection{Three-form flux}

For stabilizing the complex structure and the axio-dilaton moduli,
we  turn on type IIB three-form fluxes. The superpotential
generating the corresponding F-term scalar potential is of the familiar
Gukov-Vafa-Witten (GVW) type
\eq{
  \label{s_pot_02}
  W= \int_{\mathcal M} \bigl( F +i S\, H \bigr) \wedge \Omega_3\,
}
where $F=dC_2$ and $H=dB_2$ denote the R-R and NS-NS
three-form field strengths.
Since $W$ does not depend on the K\"ahler moduli, the scalar potential
is of no-scale type
\eq{
     V=e^K\left( G^{U\ov U} D_U W D_{\ov U}\ov W + G^{S\ov S} D_S W D_{\ov S}\ov W
     \right)
}
with Minkowski minima   at $F_U=D_U W=0$ and $F_S=D_S W=0$.
The three-form fluxes also contribute to the D3-brane tadpole
as $N_{\rm flux}=\int H\wedge {F}$ so that for compensating
the $O3$-plane tadpole one also needs localized sources in the form
of D3-branes.

It is well known that the backreaction of such a three-form flux and
of localized D3-branes on
the geometry leads to a warped CY metric \cite{Giddings:2001yu}, i.e. 
\eq{
                 ds^2=e^{2A(y)} \eta_{\mu\nu} dx^\mu dx^\nu +
                            e^{-2A(y)} \tilde g_{mn} dy^m dy^n
}
where the warp factor $A(y)$ only depends on the internal coordinates
$y$ and $\tilde g_{mn}$ denotes the Ricci-flat metric on a CY
threefold. 
Locally, the warp factor for a stack of D3-branes reads
\eq{
                 e^{-4A(y)}=1+ {4\pi g_s N\over |y|^4}\,,
}    
i.e. it blows up close to the position of the $D3$-branes.

However, a warp factor is also induced by fluxes.
Locally an $H_3$ form flux on an $A$-cycle and an
$F_3$ form flux on its symplectic dual $B$-cycle leads to the warped
metric  on the deformed  conifold. This can be described as a cone over
$T^{1,1}$ cut off  in the IR by a finite size $S^3$.  Close to the
tip,
the warp factor is related to the value of the complex structure modulus
near the conifold \cite{Candelas:1989js}  via $e^{A_{\rm con}}\sim |Z|^{1\over
  3}$. 
However, the moduli dependence is a bit more involved.
Scaling the internal metric via $\tilde g\to \lambda^2 \tilde g$
describes the breathing mode of the CY, i.e. the K\"ahler modulus 
for the overall volume. The relation is $\lambda\sim {\cal V}^{1\over
  6}$.
In \cite{Giddings:2005ff}  it was shown that the string equations of motion admit
an unconstrained deformation $\lambda$ only if the warp factor
scales non-trivially
\eq{
           e^{-4A(y)}=1+ {e^{-4A_{\rm  con}}\over \lambda^4}\sim 1+
           {1\over  ({\cal V} |Z|^2)^{2\over 3}}
\,.
}
Therefore, the warp factor can be approximated by a constant
in the so-called dilute flux limit, which in this case takes the form
\eq{
                     {\cal V} |Z|^2\gg 1\,.
}
Note that in this limit the physical size of the three-cycle $A$
\eq{
             {\rm Vol}(A)={\cal V}^{1\over 2}\left|{\textstyle \int_A \Omega_3 }\right|= ({\cal
               V} |Z|^2)^{1\over 2}
}
remains large, even with  $|Z|$ becoming small.

Let us emphasize that,  only in this limit,  one can use the usual effective
low-energy supergravity theory for the massless modes of the
CY compactification. It has been argued that in the case of
significant warping Kaluza-Klein modes  localized in the throat 
are  red-shifted  so that their mass is smaller than the mass of some of the 
stabilized former massless modes. Moreover, the derivation of a full 
effective theory for the strongly warped case has turned out to be  a
tough exercise \cite{DeWolfe:2002nn,deAlwis:2003sn,Giddings:2005ff,Shiu:2008ry} with additional  subtleties arising from mixing of the
modes and the necessity  to introduce compensator fields.

Throughout this paper we will work in the dilute flux limit and will
investigate to what extent one can achieve moduli stabilization
close to the conifold singularity, i.e. 
for small $Z$. Special emphasis is dedicated  on the question
whether new features arise that are not present
for the stabilization of  the complex structure moduli in the large complex
structure regime.

\subsection{Large volume scenario}
\label{sec_LVS}

Up to now the K\"ahler moduli remained as flat directions so that
one could simply choose the total volume of the Calabi-Yau
sufficiently large so that one stays in  the
dilute flux limit.
Since the dilute flux limit requires that we fix the total volume ${\cal V}$ at
exponentially  large values, the natural way
seems  to combine the previous framework with the large volume
scenario (LVS). 

Let us briefly recall the main aspects of the LVS.
There, one  considers a swiss-cheese
Calabi-Yau  threefold with the $\alpha'$-corrected K\"ahler potential
\eq{
                K=-2\log \left( \tau_b^{3\over 2}-\tau_s^{3\over 2}
                  +{\xi\over 2}  {\rm Re}(S)^{3\over 2} \right)\,,
}
with $\xi=-{\chi(M) \zeta(3)\over 2 (2\pi)^3}$ and where for
simplicity we are choosing a CY manifold with only two K\"ahler moduli.
In addition one has the GVW superpotential \eqref{s_pot_02} corrected
by  a non-perturbative effect like an euclidean 
D3-brane instanton or a gaugino condensate on a stack of D7-branes
\eq{
          W_{\rm LVS}(T)=W_0 +A_s\, e^{-a_s T_s}\, .
}
Here $W_0$ is the value of the  GVW-superpotential, after integrating
out the axio-dilaton and the complex structure moduli.
For the LVS minimum to exist one needs $\chi(M)<0$, i.e. $h^{2,1}(M)>h^{1,1}(M)$.
Up to CY-geometry dependent coefficients of order one, 
after freezing the axion
$\rho_s$, the relevant terms in the scalar potential read
\eq{
\label{ScalarPotLVS}
    V_{\rm LVS}(T)=e^{K_{cs}} {g_s\over 2}\Bigg({|a_s A_s|^2 \sqrt{\tau_s}\, e^{-2a_s \tau_s}\over {\cal V}} -
    {W_0\, |a_s A_s| \,\tau_s\, e^{-a_s \tau_s}\over {\cal V}^2} +{\xi\,
      W_0^2\over g_s^{3\over 2} \,{\cal V}^3}\bigg)\,.
}
Here $K_{cs}$ denotes the K\"ahler potential for the complex structure
moduli. 
In the non-supersymmetric AdS-type large-volume  minimum the K\"ahler moduli   get stabilized at
\eq{
      \tau_s={(4\xi)^{2\over 3}\over g_s} \,,\qquad
      {\cal V}={W_0\, \xi^{1\over 3} \over 2^{1\over 3}\, g_s^{1\over 2} |a_s A_s|}
        e^{a_s \tau_s}\,.
}
Note that the scalar potential $V_{\rm LVS}$ close to the LVS-minimum
is $1/{\cal V}$ suppressed relative to the former no-scale
potential for the complex structure and the axio-dilaton moduli.

The canonically normalized K\"ahler moduli masses can be computed as
the eigenvalues of the matrix $(M^2)^i{}_{j} = \frac{1}{2} K^{i
  k}\, \partial_k \partial_j V$, where the inverse of the K\"ahler metric is given at leading order in
$1/{\cal V}$ as
\eq{
	K^{\tau_b \bar{\tau}_b} &= \frac{4}{3}{\cal
          V}^{\frac{4}{3}}\,, \qquad K^{\tau_s \bar{\tau}_s} =\frac{8}{3} \sqrt{\tau_s}\, {\cal V} \\
	K^{\tau_b \bar{\tau}_s} &= K^{\tau_s \bar{\tau}_b} = 4 \tau_s\, {\cal V}^{\frac{2}{3}}\,.
	}
At leading order, the masses of the K\"ahler moduli are
\eq{
\label{lvs_mass}
                M^2_{\tau_b}&\sim O(1)  {W_0^2\, \xi\over g_s^{1\over
                    2}\, {\cal V}^3} M_{\rm pl}^2\,, \qquad M_{\rho_b}^2 \sim 0\,,\\
               M^2_{\tau_s}&\sim  M^2_{\rho_s}\sim O(1) {a_s^2\, W_0^2\,  \xi^{4\over
                   3}\over g_s\, {\cal V}^2} M_{\rm pl}^2 \, .
}
For later purposes, note that the masses do not depend on the parameter $A_s$.

\section{Moduli stabilization close to  the conifold}
\label{sec:ModStab}

In this section we first provide the form of the periods for the
quintic
in the vicinity of the conifold singularity.
This serves as the prototype example for the subsequent analysis 
of dynamical moduli stabilization. In our analysis we restrict
ourselves to the three (final) dynamical moduli: axio-dilaton, 
the complex structure governing the size of the three-cycle vanishing
at the conifold locus and the K\"ahler moduli.

\subsection{Periods of the quintic}

As a concrete example of a Calabi-Yau  manifold featuring a conifold singularity,
we consider the mirror dual of the quintic $\mathbb P_4[5]^{(101,1)}$,
whose single complex structure modulus is given by the complex
parameter $\psi$ in the hypersurface constraint
\eq{   
P=\sum_{i=1}^5  Z_i^5 - 5\psi \prod_{i=1}^5  Z_i =0\,.
}       
For the co-dimension one locus $\psi=1$, this hypersurfaces becomes singular, i.e.
$P=\partial_i P=0$ for $i=1,\ldots,5$.
The second derivatives do not vanish so that one has a conifold
singularity. It is known that at the singularity $u=5(\psi-1)=0$,
a three-cycle $B^1$ shrinks to zero size, i.e. the corresponding
period has to vanish like $F_1=\int_{B^1} \Omega_3\sim u +O(u^2)$. 
Moreover, for a closed loop around the conifold singularity, 
the symplectic dual period undergoes a monodromy
$X^1\to X^1 + F_1$. The remaining periods should stay  finite 
at the conifold locus.

For the concrete model of  the mirror of the quintic, in the regime
$|\psi|<1$ a basis of periods solving the Picard-Fuchs equation can
be derived from the fundamental period \cite{Candelas:1990rm}
\eq{
           \varpi_f(\psi)=-{1\over 5}\sum_{n=1}^\infty {\lambda^{2n}\,
             \Gamma\left({n\over 5}\right)
  (5\,\psi)^n \over
    \Gamma(n)\, \Gamma^4\left(1-{n\over 5}\right)}\,.
}
via
\eq{
       \varpi_i(\psi)=-\left(  {2\pi i\over 5}\right)^3 \varpi_f(\lambda^i \psi )
}
with $i=0,1,2,4$ and $\lambda=\exp(2\pi i/5)$.
These do not yet form a symplectic basis, which was determined 
explicitly in \cite{Candelas:1990rm}. 
When expanded around the conifold locus $u\sim 0$ and after an Sp$(4; \mathbb{Z})$ transformation, they
take the form \cite{Curio:2000sc,Huang:2006hq}
\eq{
      F_0&= \tilde a_0 +\tilde b_0\, u +\ldots \,,\\[0.1cm]
      F_1&=a u+\ldots \\[0.1cm]
       X^0&=a^0+b^0\, u + \ldots \,,\\[0.1cm]
        X^1&=-{1\over
        2\pi i} F_1 \log u + c +d\, u+\ldots
}
with the following numerical values  of the parameters
\eq{
\label{paraquintic}
    a&={\sqrt{5}\over 2\pi i}\,, \quad c=1.07072586843016
      \,, \quad d=-0.0247076138044847\\
    a^0&=12.3900325542991 \,, \quad  b^0=2.033209433405164 \\
    \tilde a_0&=6.19501627714957-0.64678699225205\,i\\
     \tilde b_0&=1.016604716702582-0.075383347561773\,i\,.
}
Half of the periods can be considered as homogeneous coordinates on
the complex structure moduli space. Now we introduce inhomogeneous coordinates by dividing by the period $X^0$.
Introducing the  new complex structure modulus  
\eq{
        Z={F_1\over X^0}={a\over a^0} u +O(u^2)\,,
}
which in the following we also call the ``conic'' modulus,
the period vector $\Pi^T=(F_0,F_1,X^0,X^1)$ can  be expressed as
\eq{
\label{periods}
    \Pi=X^0\,\left( \begin{matrix}  \tilde A_0 -\tilde B_0 Z
               +O(Z^2) \\ Z\\ 1\\
                            -  {1\over 2\pi i} Z\log Z + C + DZ + O(Z^2) \end{matrix}   \right)
}
with  parameters
\eq{
     \tilde A_0&={\tilde a_0\over a^0}={\textstyle {1\over 2}} -0.05220220281242\, i\\
     \tilde B_0 &= - { a^0 \tilde b_0 - b^0 \tilde a_0\over a\,
       a^0}= 0.08641832567733 \\
     C &= {c\over a^0}= 0.08641832567733 =\tilde B_0\\
       D&={1\over a}\left(d-{b^0 c\over a^0}+{a\over 2\pi
           i}\log\left({a\over a^0}\right)\right)=- \frac{1}{4} + 0.00185911259239\, i\,.
}  
Note the non-trivial relation $C=\tilde B_0$.
For these values, the corresponding K\"ahler potential for the complex structure modulus
is given by
\eq{
\label{kahlerconi}
           K_{\rm cs}&=-\log\left[ -i \Pi^\dagger \Sigma\, \Pi \right]\\[0.1cm]
    &=-\log\left[ {1\over 2\pi} |Z|^2\, \log( |Z|^2) + A + O(|Z|^2)\right]
}
with  $A=0.10440$ and the symplectic pairing
\eq{
       \Sigma=\left(\begin{matrix} 0 & 0 & 1 & 0\\0 & 0 & 0 & 1\\
                     -1 & 0 & 0 & 0\\  0 & -1 & 0 &
                     0\end{matrix}\right)\,.
}
Note that the linear terms in $K_{\rm cs}$
cancel\footnote{Note that the cancellation only appears after a proper $Sp(4;\mathbb{Z})$ transformation, thus our setup is in agreement with results e.g. from \cite{Bizet:2016paj}.}
 so that, as advocated in \cite{Garcia-Etxebarria:2014wla},
 the K\"ahler potential respects the continuous shift symmetry
$Z\to e^{i\theta} Z$.
In section \ref{period_b} we generalize these results even up to
second order to the
mirror of the Calabi-Yau threefold $\mathbb P_{11226}[12]^{(128,2)}$ that
features two complex structure moduli. 

Throughout this paper we consider  these two Calabi-Yaus as 
prime examples of   complex structure moduli spaces developing
a conifold singularity. However, when analyzing concrete models
of moduli stabilization, 
we allow ourselves some more flexibility with respect to  the concrete values of the 
numerical parameters in the periods.

\subsection{Flux induced exponential mass hierarchies}
\label{sec_fluxexp}

Let us now discuss whether the
GVW-superpotential admits vacua in which the complex structure (conic)
modulus $Z$ is  dynamically 
stabilized close to the conifold singularity $Z=0$. The first concrete
proposal of this has been made in \cite{Giddings:2001yu}, which we also take as
our starting point.

\subsubsection*{Stabilizing the  conic  modulus}

Consider a CY  like the mirror of the  quintic that develops a
conifold singularity at $Z=0$ and which has  four periods
that around the conifold locus $Z\sim 0$ admit an expansion \eqref{periods}
\begin{align}
       X^0&= 1, &    X^1&= - \frac{1}{2 \pi i} Z \log Z + \tilde B_0 + D Z + \ldots \\
      F_0&= \tilde A_0 -\tilde B_0\, Z +\ldots \,, &
        F_1&= Z \, . \nonumber
\end{align}
Now we freeze the complex structure modulus $Z$ by turning on
three-form fluxes so that the superpotential reads
\begin{align}
\label{superpotexa}
      W&=f\, X^1 + i h S F_1 + i h' S F_0\\
      &=f\left(- { 1\over 2\pi i} Z \log Z + \tilde B_0 +D \, Z + \ldots \right) + i h S\,
       Z + ih' S (\tilde A_0-\tilde B_0\,Z+...) \, . \nonumber
\end{align}
In the follwoing we take $h \gg h'\,\tilde B_0 $.
Let us first freeze the complex structure modulus via $F_Z=0$, which
up to terms that vanish in the $Z\to 0$ limit leads to
\eq{
\label{ftermrela}
        \underbrace{ {f \over 2\pi i}  \log Z - ih\,  S}_{{\rm
            order}\;\log Z} \, \underbrace{ + \, {f\over 2\pi i} - D \, f
        }_{{\rm order}\, O(1)}+\ldots=0 \, .
}
Note that this  contribution entirely comes from the $\partial_Z W$
term in $F_Z$ and that the contribution from  $(\partial_Z  K) W$ in
$F_Z$ is subleading\footnote{While the present work was almost
  finished (see \cite{BHG}), this order one correction in
  \eqref{ftermrela} has also been observed in \cite{Bizet:2016paj}.}.
Moreover, we assume that, after all, the axio-dilaton can be stabilized
such that $h/(f g_s)\gg 1$ and that $h\gg h'$.
Therefore at leading order one finds 
\eq{
\label{stableZ}
                    Z\sim \hat C\, e^{-{2\pi h\over f} S}\,,\qquad{\rm
                      with}\quad
                      \hat C=\exp\left( {\textstyle -1+2 \pi i \,D}\right)
}
so that for a sufficiently large exponent one can indeed stabilize the
complex structure modulus close to the conifold singularity\footnote{For the
quintic one finds $\hat C=\exp(-1.01-1.57 \, i)$.}.

After invoking the relation \eqref{ftermrela}, the second $F$-term $F_S$ yields 
\eq{
 D_S W=\left(i\,h\hat C e^{-{2\pi h\over f} S} + i h'\,\tilde A_0\right)
   - {\textstyle {1\over S+\ov S}} \left(  \tilde B_0 \, f + {f\over 2\pi i} \hat C\, e^{-{2\pi h\over f}S}
      + i h' S \tilde A_0\right) .
}
We observe that this F-term  is identical to the one that can be derived  by
inserting the solution \eqref{stableZ} directly into the superpotential \eqref{superpotexa} to obtain an
effective superpotential for the axio-dilaton modulus
\eq{
\label{superpoteff}
     W_{\rm eff}= \tilde B_0 \,f + {f\over 2\pi i} \hat C\, e^{-{2\pi h\over f}S}
     + i h' S \tilde A_0+\ldots
}
This superpotential features the two 
 striking features
\begin{itemize}
\item{Besides flux induced polynomial terms
like $ih' S$ the effective superpotential contains also infinitely many
exponentially suppressed terms that, though
also flux induced, have the same form as coming from some $D(-1)$
instantons. }
\item{The exponential term suggests that the continuous shift
symmetry $S\to S+i\theta$ is broken to a discrete one by the
non-vanishing $h$-flux. This is evident from the superpotential
\eqref{superpotexa}, as  a discrete shift of the universal axion
can be compensated by changing the branch of the $\log Z$-term in $W$.
}
\end{itemize}
Therefore, the effective superpotential \eqref{superpoteff} offers
the possibility  to mimic the behavior of
non-perturbative effects via flux induced tree-level contributions to
$W$. Before we employ  \eqref{superpoteff} to stabilize the
axio-dilaton such that  $h/(f g_s)\gg 1$, let us compute the 
mass of the conic modulus $Z$.

\subsubsection*{Mass of conic  modulus}

Since the complex structure modulus $Z$ is fixed via
$D_Z W=0$, for determining its mass we can evaluate
\eq{
 V_{Z\ov Z}=\partial_Z\partial_{\ov Z} V=e^K\,  G^{Z\ov Z} \partial_Z(D_Z
 W) \,\partial_{\ov Z} (D_{\ov Z}\ov W)\Big\vert_{D_Z W=0}\,.
}
In our case $D_Z W$ is holomorphic at leading order and hence the masses
of the two real scalars in $Z$ are degenerate.
For the metric one obtains at leading order 
\eq{
       G_{Z\ov Z}\sim -{1\over 2\pi A} \log( |Z|^2)\,
}
so that
\eq{
     V_{Z\ov Z}\sim -{1\over 2 {\rm Re(S)} {\cal V}^2 |Z|^2}
     {f^2\over 2\pi \log(|Z|^2)} \, .
}
With $M^2_{Z}={1\over 2}G^{Z\ov Z} V_{Z\ov Z} $, the mass of the canonically
normalized complex structure modulus becomes
\eq{
\label{masscomplexstructure}
     M^2_{Z}\sim {M_{\rm pl}^2\over 4 {\rm Re(S)} {\cal V}^2 |Z|^2 }
     {A f^2\over  \log^2(|Z|^2)} 
}
Using the expression for the string scale, $M_{\rm s}^2={M_{\rm pl}^2
  g_s^{1\over 2}\over {\cal V} }$, one can write
\eq{
     M^2_{Z}\sim {M_{\rm s}^2\over {\cal V} |Z|^2 }
     {A f^4 g_s^{5\over 2}\over  16 \pi^2 h^2  } \,.
}
Due to the $|Z|^2\sim \exp(-{4\pi h\over f g_s})$ factor in the
denominator this mass is exponentially enhanced so that it only makes
sense for ${\cal V}|Z|^2\gg 1$, i.e. for exponentially large
volume. 
To summarize: 
\begin{itemize}
\item{The mass of the conic modulus $Z$ comes out exponentially
enlarged so that the volume eventually has to be chosen/frozen 
at exponentially large values. This makes the large volume scenario
the natural candidate for K\"ahler-moduli stabilization.
}
\item{The used effective supergravity theory by itself indicates its limitation, i.e.
that it  is 
applicable only in the dilute flux regime ${\cal V}|Z|^2\gg 1$ where warping
is negligible.}
\end{itemize} 

\subsubsection*{Mass of axio-dilaton}

For stabilizing the axio-dilaton and computing its mass, let us  distinguish 
the two cases whether the flux $h'$ vanishes or not.

\vspace{0.5cm}
\noindent
{\it Case A: $h'\ne 0$}

\vspace{0.2cm}
In this case the exponential term in the effective superpotential 
\eqref{superpoteff} can be neglected against the linear term so
that
\eq{
     W_{\rm eff}= \tilde B_0\,f + i h' S \tilde A_0
}
and
\eq{
 D_S W_{\rm eff}= - {1\over S+\ov S} \left(-i h' \ov S \tilde A_0 + \tilde B_0 \,f \right)\,.
}
One gets
the stabilized axio-dilaton
\eq{
          {1\over g_s}=  {f\over h'} \,{\rm Im}\Big({\tilde B_0 \over \tilde A_0} \Big)\,,\qquad C_0= - {f\over h'}\,{\rm Re}\Big({\tilde B_0 \over \tilde A_0} \Big)\,.
}
For the quintic one finds ${\rm Im}({\tilde B_0 \over \tilde A_0}) \sim 0.018$ so that we should
choose $f, h'>0$.
For the complex structure modulus we thus obtain
\eq{
 Z \sim \hat C\, \exp\left( - 2 \pi i \,{\ov{ \tilde B_0} \over \ov{ \tilde A_0}} {h\over  h'}\right)\, . 
}
For fixing $g_s<1$ in the perturbative regime and to have $Z$ close
to the conifold we require $|h|>|h'|$ and $|f|>|h'|$, while the relative signs of the fluxes depend on the sign of ${\rm Im}({\tilde B_0 \over \tilde A_0})$. Note that there
is no restriction on the the relative size of $h$ and $f$.

The degenerate mass of the complex axio-dilaton can be determined straightforwardly as
\eq{
\label{mass_dila}
       M^2_{S}={M_{\rm pl}^2\over {\cal V}^2 {\rm Re}(S)}{|\tilde B_0 f|^2 \over A}
}
so that one obtains
\eq{
           {M^2_S\over M_Z^2}\sim |Z|^2 \sim \exp\left(- 4\pi \,{\rm Im} {\Big({\tilde B_0 \over \tilde A_0}\Big)} {h\over  h'}\right)\,,
}
i.e. the mass of the axio-dilaton is exponentially suppressed against
the mass of the complex structure modulus. Note that this a posteriori
justifies the use of the effective superpotential \eqref{superpoteff} for the axio-dilaton modulus.
Let us emphasize that  this effect is not due to  warping, as it remains in the
dilute flux limit. In the strongly warped case it is usually the
red-shifted modes in the throat that become light, whether here it is 
actually a bulk closed string mode.
 
\vspace{0.5cm}
\noindent
{\it Case B: $h' = 0$}

\vspace{0.2cm}
In this case the effective superpotential is of  KKLT-type
\eq{
\label{superpoteffb}
     W_{\rm eff}= \tilde{B}_0 \, f + {f\over 2\pi i} \hat C\, e^{-{2\pi h\over f}S} \, .
}
In the large ${\rm Re}(S) = g_s^{-1}$ regime that we are working in, the resulting
F-term condition $F_S=0$ can be written as
\eq{
       i h \hat C e^{-{2\pi h\over f}S} - {f \tilde{B}_0\over 2{\rm Re(S)}} + \ldots =0 \, .
}
Abbreviating $\tilde s:={2\pi h\over f}{\rm Re}(S)$ and taking the
values of the coefficients for the quintic \eqref{paraquintic} one
gets  for the axion $C_0= - \frac{f}{h} \left( {\rm Re} (D) + \frac{1}{4} \right)$ and a
transcendental relation for the saxion
\eq{        e^{\tilde s}= { |\hat C| \over \pi \tilde{B}_0} \tilde s\,,\qquad\qquad {\rm
    with}\quad \hat C=\exp\left({\textstyle -1+{2\pi i D}}\right)
}
where the prefactor does not depend on the fluxes but only
on the coefficients appearing in the periods of the underlying
CY threefold.

%
%
For the quintic one obtains for the prefactor  $\lambda={|\hat C| \over \pi \tilde{B}_0}
\sim 1.34$, a value that is way too low to admit solutions to this
equation in the regime $\tilde s>1$.
Let us now assume that we have a threefold where $\lambda\gg 1$.
In that case, one would find a KKLT-like minimum with $\tilde s>1$ as the solution of
the transcendental equation $\log \tilde s=\tilde s -\log \lambda$.

In the framework of large field inflation via axions it has been
claimed that in the controllable regime the axion decay constant
cannot be larger than one (in natural units where  $M_{\rm pl}=1$).
Since here we generate the exponential not directly via instantons
but via moduli stabilization close to the conifold, it is interesting
to investigate the appearing value of the axion decay constant.
For that purpose, let us consider the scalar potential for the axion $\theta=C_0$ after
plugging in the minimum value of $\tilde s$
(for simplicity we set ${\rm Re} (D) = 0$)
\eq{
         V(\theta)={1\over A\, {\cal V}^2 (S + \ov S)} \, 2 |f \tilde B_0|^2\left[
           1-\cos\left({\textstyle {2\pi h\over f} \theta}\right) \right]\,.
  }
To compute the axion decay constant one has to go to the canonically
normalized field $\tilde\theta=\theta/(\sqrt 2 {\rm Re}(S))$. This yields
\eq{
\label{effpotaxion}
           V(\tilde\theta)={1\over A\, {\cal V}^2 {\rm Re}(S)} \, |f \tilde B_0|^2\left[
           1-\cos\left({\textstyle \sqrt{2} \tilde s\, \tilde \theta}\right) \right]
  }
so that $f_{\tilde\theta}=1/(\sqrt{2} \tilde s)<1$ in the regime where we have
control. 

From \eqref{effpotaxion} one can directly read off  the mass of the
axion
\eq{    M_{\tilde \theta}^2\sim  {M_{\rm pl}^2 h f \over {\cal V}^2} {4 \pi
  |\tilde B_0|^2 \tilde s\over A}
}
For this KKLT-like minimum of the axio-dilaton 
the mass of the complex structure modulus \eqref{masscomplexstructure}
can be simplified to
\eq{
         M_Z^2\sim {M_{\rm pl}^2  h f \over {\cal V}^2} {A\over 8 \pi \, |\tilde B_0|^2
           \tilde s}\,.
}
Recall that the value of $\tilde s$ only depends on the parameter $\lambda$.
Therefore, up to CY threefold dependent data, the two masses
scale in the same way with the fluxes and the overall volume.
Therefore, in this case there is no exponential hierarchy among them.

\subsection{Conic LVS scenario}
\label{sec_conLVS}

So far, the scalar potential did not
depend on the K\"ahler moduli, i.e. in particular on the overall
volume of the CY.  As we have seen, for the mass of the conic modulus $Z$
to remain below the string scale, we need ${\cal V} |Z|^2 \gg 1$.
Therefore, one needs to dynamically freeze the overall volume modulus
at an exponentially large size. This makes it natural to combine
our approach with the large volume scenario, reviewed in section \ref{sec_LVS}.

\subsubsection*{K\"ahler moduli stabilization}

Note that for the LVS minimum to exist one needs $h^{21}>h^{11}$,
which is clearly not satisfied for the mirror of the quintic. The
minimal setup would therefore be a CY with Hodge numbers
$(h^{21},h^{11})=(3,2)$. Close to a conifold singularity with
$|Z|\ll1$ we expect that the two additional complex structure moduli
can be stabilized via fluxes at the  mass scale of the axio-dilaton \eqref{mass_dila}.

Integrating out all these moduli, we finally have to treat the issue
of K\"ahler moduli stabilization. 
Since we do not explicitly know the one-loop Pfaffian, we allow
it to depend polynomially on $Z$ and make the ansatz 
\eq{
          W_{\rm inst}(T_s)=W_0 +A_s\, Z^N\, e^{- a_s T_s}\, 
}
with  $W_0\sim   f $ and the c-number
$ Z\sim \hat C\, \exp\big(-{2\pi h\over f} S\big)$. Now we proceed
as in section \ref{sec_LVS}, where only the value of $A_s$ has been
changed according to $A_s\to A_s Z^N$ and is now an exponentially small number.
In the non-supersymmetric AdS-type large-volume  minimum the K\"ahler
moduli   get now stabilized at
\eq{
      \tau_s={(4\xi)^{2\over 3}\over g_s} \,,\qquad
      {\cal V}={W_0\, \xi^{1\over 3} \over 2^{1\over 3}\, g_s^{1\over
          2} |a_s A_s Z^N|}
        e^{a_s \tau_s}\,.
}
For $N>0$ the exponentially small value of $Z$ only further enhances the size of
${\cal V}$. For the crucial combination we thus obtain
\eq{
     {\cal V} |Z|^2\sim \exp\left[ {a_s\over g_s} \left(
     {h(N-2)\over f} + (4\xi)^{2\over 3}\right)\right] \, .
}
For $N\ge 2$ this is naturally larger than one and for $N\le 1$ we can
choose $f\gg h$ so that the exponent
becomes positive. Recall from section \ref{sec_fluxexp} that there was no direct
correlation between the fluxes $f$ and $h$. For staying in the
perturbative regime only $h>h'$ and $f>h'$ was required.

\subsubsection*{Mass hierarchy}

Let us now consider the moduli masses at the minimum. We observe that the
masses of the K\"ahler moduli in \eqref{lvs_mass} do not depend on the
parameter $A_s$ and therefore not on $Z$. 
Thus, one still finds
\eq{
                M^2_{\tau_b}&\sim {W_0^2\, \xi\over A\, g_s^{1\over
                    2}\, {\cal V}^3} M_{\rm pl}^2\,, \qquad M_{\rho_b}^2 \sim 0\,,\\
               M^2_{\tau_s}&\sim  M^2_{\rho_s}\sim {a_s^2\, W_0^2\,  \xi^{4\over
                   3}\over A\, g_s\, {\cal V}^2} M_{\rm pl}^2 \, .
}
The gravitino mass is given by 
\eq{
        M^2_{3/2} = e^K |W|^2\sim \frac{g_s\, W_0^2}{A {\cal V}^2} \, M^2_{{\rm Pl}}\,.
}
In Table \ref{table_LVSInstMasses} we summarize  all relevant mass
scales, where we only displayed the dependence on ${\cal V},g_s$,
$Z$ and the large flux $f$. We  choose $h' \neq 0$ to determine the axio-dilaton
     mass $M_S$ and ordered the mass scales in the perturbative large ${\cal V}$
     and small $g_s,|Z|$ regime.
\begin{table}[ht]
\centering
\renewcommand{\arraystretch}{2.2}
\begin{tabular}{|c|c|}
  \hline
   Scale & $(\text{Mass})^2$ in $M^2_{\rm Pl}$ \\
  \hline\hline
  string scale $M^2_{\rm s}$ & $\displaystyle \frac{g_s^{1/2}}{{\cal V}}$ \\
  Kaluza-Klein scale $M^2_{\rm KK}$ & $\displaystyle \frac{1}{{\cal V}^{4/3}}$ \\
  conic  modulus
  $M^2_Z$ & $\displaystyle \frac{f^4 g_s^{3}}{{\cal V}^2 |Z|^2}$ \\
  small K\"ahler modulus $M^2_{\tau_s}$ &
  $\displaystyle \frac{f^2 }{g_s \mathcal{V}^2}$ \\
  gravitino $M^2_{3/2}$ & $\displaystyle \frac{f^2\, g_s}{{\cal V}^2}$ \\
  axio-dilaton $M^2_S$ and c.s. moduli  &$\displaystyle \frac{f^2 g_s}{{\cal V}^2}$\\
  large K\"ahler modulus $M^2_{\tau_b}$ &
  $\displaystyle \frac{ f^2 }{g_s^{1/2}\, \mathcal{V}^3 }$  \\[4mm]
\hline
     \end{tabular} 
     \caption{\small Mass scales and moduli masses for the conic LVS. }
      \label{table_LVSInstMasses}
\end{table}\\

\noindent
Hence,  we obtain the mass hierarchy
\eq{
	M_{\tau_b} \, < \, M_{\tau_s} \, \sim \, M_{S} \, < \, M_{Z} < \, M_{\rm KK}\, <
        \, M_{\rm s}\, < \, M_{\rm Pl} \, .
	}
Since the mass of the modulus $\tau_s$ is of the same volume order
$1/\mathcal{V}^2$ as the axio-dilaton $S$, one might worry whether
integrating out $S$ and only considering the K\"ahler moduli is actually
justified. However, as shown in \cite{Balasubramanian:2005zx} the
minimum above remains a minimum of the full potential including
axio-dilaton and complex structure moduli. The reason is that these
moduli are flux-stabilized and enter the scalar potential already at the order
$\mathcal{O} (\frac{1}{\mathcal{V}^2})$ that, 
in the large volume regime, dominate over
the  terms of  order $\mathcal{O} (\frac{1}{\mathcal{V}^3})$
by which the K\"ahler moduli are stabilized.

We conclude that flux stabilization of the complex structure and the
axio-dilaton close to the conifold singularity
can be consistently combined with the LVS scenario, thus  guaranteeing 
a reliable effective field theory approach where warping is
negligible.  The masses of the moduli get split up so that one
gains parametric control over their ratios.

\section{Towards axion alignment}

In this section we generalize the previous set-up to models
with more complex structure moduli. The motivation is twofold.
First, we would like to see whether the periods of threefolds
with more complex structure moduli still arrange themselves 
to provide K\"ahler potentials with shift-symmetries.
Second, with more complex structure moduli we will be able
to address recent questions about large field inflation.
Concretely, we will design a model of aligned inflation according to the proposal of
\cite{Hebecker:2015rya}. There one linear combination of two axions was stabilized via fluxes by
polynomial terms in $W$, while the orthogonal direction was stabilized
by exponentially subleading terms in $W$.  
This provides another possible  application of the
exponential terms in the effective superpotential after integrating
out the most heavy complex structure modulus $Z$.
Models with the inflaton being identified with an axion-like  complex
structure modulus
have also been investigated in \cite{Blumenhagen:2014nba,Hebecker:2014kva,Kobayashi:2015aaa,Bizet:2016paj}.

\subsection{Periods for $\mathbb{P}_{11226}[12]$}
\label{period_b}

Similar to the quintic,  we now numerically evaluate the periods of the Calabi-Yau 
threefold $\mathbb{P}_{11226}[12]^{(128,2)}$ in the vicinity of the
conifold locus. We are particularly interested in computing
the K\"ahler potential to see whether it exhibits any shift-symmetry.

The mirror of the above  manifold has two complex structure moduli that
appear as deformations of the hypersurface constraint
\eq{
          P=z_1^{12}+z_2^{12}+z_3^{6}+z_4^{6}+z_5^{2}-12\psi\,
           z_1\,z_2\,z_3\,z_4\,z_5-2\phi\, z^6_1\,z^6_2\,.
}
The codimension one conifold locus is at $864\,\psi^6+\phi=1$.
In the regime of small $\psi$ and $\phi$ the fundamental period is
given by \cite{Berglund:1993ax,Candelas:1993dm}
\eq{
\label{periodfund}
\varpi_f(\psi,\phi)=-{1\over 6}\sum_{n=1}^\infty {\Gamma\left({n\over 6}\right)
  (-12\,\psi)^n \; u_{-{n\over 6}}(\phi)\over
    \Gamma(n)\, \Gamma^2\left(1-{n\over
        6}\right)\,\Gamma\left(1-{n\over 2}\right)}
}
with
\eq{    u_{-{n\over 6}}(\phi)={e^{-i\pi{n\over 12}} \over
    2\,\Gamma\left({n\over 6}\right)}
       \sum_{m=0}^\infty {e^{i\pi{m\over 2}}\, \Gamma\left({m\over
             2}+{n\over 12}\right)  \, (2\phi)^m\over m!\, 
  \Gamma\left(1-{m\over 2}-{n\over 12}\right) }\,.
}
These expressions are  valid for  $|\phi|<1$ and $\left|{864\psi^6\over \phi\pm1}\right|<1$.
A basis of periods solving the Picard-Fuchs equations can be derived
from $\varpi_f$ via
\eq{
       \varpi_i(\psi,\phi)=-{(2\pi i)^3\over \psi} \varpi_f(\lambda^i \psi,
       \lambda^{6i} \phi)
}
with $i=0,\dots,5$ and $\lambda=\exp(\pi i/6)$.
These periods do  not yet form a symplectic basis. The latter was
determined in \cite{Kaste:1999id,Giryavets:2003vd} as
\eq{
      \left(\begin{matrix}   F_0 \\ F_1 \\ F_2\\X^0\\X^1\\X^2\end{matrix}\right)=
      \left(\begin{matrix}   
                                           {3\over 2} & {3\over 2} &
                                           {1\over 2} & {1\over 2} &
                                           -{1\over 2} & -{1\over 2}\\
                                        -1 & 1 & 0 & 0 & 0 & 0\\
                                         1 & 0 & 1 & 0 & 0 & 0\\
                                          -{1\over 2} & 0 &  {1\over
                                            2} & 0 & {1\over 2} & 0\\
                                        1 & 0 & 0 & 0 & 0 & 0\\
                                           {1\over 2} & {1\over 2} &
                                           -{1\over 2} & {1\over 2} &
                                           -{1\over 2} & {1\over
                                             2}\end{matrix}\right)\,
    \left(\begin{matrix}   \varpi_0 \\ \varpi_1 \\ \varpi_2\\\varpi_3\\\varpi_4\\\varpi_5\end{matrix}\right)\,.
}

We now want to evaluate these periods close to the conifold locus,
which lies on the boundary of validity of the expansion \eqref{periodfund}.
Concretely,  we restrict to the region around the point
$\psi=\psi_0=864^{-{1\over 6}}$ and $\phi=0$ on the conifold locus.
Writing $\psi=\psi_0+\xi$, we numerically evaluate the periods
up to quadratic order in $(\xi,\phi)$. This has also been done up to
linear order in
\cite{Conlon:2004ds}, but not to the level of accuracy that we need for our purposes.
The numerical computation of the sum in  \eqref{periodfund} up to
$n\sim 20000$  gives \footnote{For some of these numbers we gained a better precision, which was then 
also used in the following computations.}

\begin{eqnarray}
    &&\hspace{-0.6cm}F_0=4323.04i - 1548.4i\, \xi+107.7i\,\phi
  -3893.22i\, \xi^2-278.46i\,\xi\phi +27.78i\, \phi^2\,,\nonumber\\[0.1cm]
     &&\hspace{-0.6cm}F_1=3191\, \xi+172.29\,\phi 
   +7583.59\,\xi^2-533.36\, \xi\phi+65.2\,\phi^2\,,\\[0.1cm]
    &&\hspace{-0.6cm}F_2=(-492.72+1976.76i)+(372.45-302.3i)\xi-(258.97 +58.87i)\phi \nonumber\\
   &&\hspace{-0.6cm}\quad\quad\  -(436.95 - 262.39i)\xi^2-( 6.5 + 14.41i)
   \xi\phi-(3.09 - 24.14i) \phi^2
   \nonumber
\end{eqnarray}
and
\begin{eqnarray}
    &&\hspace{-0.6cm} X^0=-(994.58+184.76i)+(859.48+471.9i)
    \xi+(10.04-112.7i)\phi \nonumber\\
 &&\hspace{-0.6cm}\quad\quad+(1831.84 + 2209i)\xi^2-( 136.09 - 124.82i)\xi\phi+(3.13 + 10.25i)\phi^2\,,\nonumber\\
   &&\hspace{-0.6cm} X^1=-{1\over 2\pi i} F_1\log F_1 +784.36i - 4997.53i\,\xi- 185.8i\,\phi+\ldots\,,\\
  &&\hspace{-0.6cm} X^2=369.52i-943.81i\,\xi+225.4i\,\phi 
      -4418i\,\xi^2-249.64i\,\xi\phi - 20.5i\, \phi^2\, .  \nonumber
\end{eqnarray}
Note that the order two terms in $X^1$ will not be relevant for computing
the K\"ahler potential up to quadratic order.
Next, we go to inhomogeneous coordinates where $F_0=1$ and 
substitute 
\eq{
\phi\to -18.52\,\xi + 25.09i\,Z - 231.17\, \xi^2 +  408.85i\,\xi Z + 222.58\, Z^2
}
followed by a second substitution
\eq{
\xi\to 1.97\, Y+1.13i\,Z -62.84\, Y^2 -8.7i\, ZY+4.07\, Z^2\,.
}

Finally, in terms of the fields $Z$ and $Y$, up to quadratic order, the periods take the form
\eq{
       F_0&=1\,,\\
       F_1&=Z\,,\\
     F_2&=(0.46 + 0.11i) + (1.10-2.17i)Y  -0.19\,Z \\
     & -(7.34 - 14.73i)\, Y^2 +(2.71 + 1.42i)\, YZ +(0.11 - 1.69i)\,Z^2
}
and
\eq{
X^0&=(-0.04 + 0.23i)+( 1.10 +0.06i) Y + 0.17\,Z \\
&  -(7.34 + 1.83i)\,Y^2 + (0.55 + 1.42i)\,YZ + (0.11 - 0.17i)\,Z^2\,,\\
X^1&=-{1\over 2\pi i}Z\log Z + 0.18 - 0.42\,Y -  1.43 i\, Z +\ldots\,,\\[0.1cm]
X^2&=0.09 - 2.19\,Y + 14.67\,Y^2 - 2.84i\,YZ - 0.22\,Z^2\,.
} 
Here we have set  a numerical number to 
zero, if it vanishes up to the order $O(10^{-4})$.
Moreover, the numbers appearing in the periods are not unrelated.
Up to order $O(10^{-4})$, we find e.g.
\begin{eqnarray}
         2X^0+X^2\!\!\!&=&\!\!\!0.46i + 0.13i\, Y +0.33\, Z - 3.66i\, Y^2
  +1.11\,YZ-0.34i\, Z^2\\[0.1cm]
          F_2-X^0\!\!\!&=&\!\!\!(0.5 - 0.11i) -2.24i Y -0.36\, Z
    +16.56i\, Y^2+2.16\, YZ -1.52i\, Z^2\,.\nonumber
\end{eqnarray}
Due to these relations, the resulting K\"ahler potential at linear order simplifies considerably 
\eq{
K_{\rm cs}&=-\log\left[- i \Pi^\dagger \Sigma\, \Pi \right]\\[0.1cm]
&=-\log\left[{1\over 2\pi} |Z|^2 \log\left( |Z|^2\right) +A +
{\rm  Re}Y+    B \,({\rm  Re}Y)^2 +C\, |Z|^2\ldots\right]
}
with $A=0.44$ and $B=- 19.05$ and $C= -2.86$. 
As for the quintic, the K\"ahler potential exhibits the shift symmetry
$Z\to e^{i\theta}Z$ and in addition the shift symmetry ${\rm Im}(Y)\to
{\rm Im}(Y) +\theta$. Therefore, in this regime close to the conifold,
${\rm Im}(Y)$ behaves like an axion.

\subsection{Freezing axio-dilaton and complex structures }

Motivated by the previous periods and the shift-symmetric K\"ahler
potential  we analyze the resulting supergravity models
for moduli stabilization. Here we proceed analogously to
section 3 and first stabilize the $Z$ modulus by the usual combination
of  R-R and NS-NS fluxes on the two cycles related to the conifold
singularity.
Integrating out this most heavy complex structure modulus
leads to  effective K\"ahler- and superpotentials that we
subsequently study in a more phenomenological manner, i.e.
without reference to a concrete CY threefold.  Thus, we
allow us to be more flexible with independent fluxes
and numerical prefactors than  an actual threefold example
may permit.

For such an  effective model with two complex structure moduli $(Z,Y)$,
up to linear order in  $Y$, we  make the ansatz
\eq{
K_{\rm eff}= - 2 \log \mathcal V  -\log(S+\ov S) -\log\left(A+{\textstyle {1\over 2}}(Y+\ov Y)\right)
}
and
\eq{
            W_{\rm eff}=f\, \alpha + h'\, \beta  S +\hat f'\, \gamma Y
} 
with $\alpha,\beta,\gamma\in \mathbb C$ and $A\in\mathbb R$. 
However, the minimum conditions $D_S W=D_YW=0$ 
only admit solutions with ${\rm Re}(S)=0$, i.e. in the unphysical
domain. Therefore, our ansatz is not yet sufficiently generic.

Adding the second order term  to the K\"ahler potential,
we make the ansatz
\eq{
\label{effaxionmodelb}
K_{\rm eff}&= - 2 \log \mathcal V -\log(S+\ov S) -\log\Big(A+\kappa\, {\rm Re}Y -({\rm Re}Y)^2\Big)\\[0.2cm]
         W^{(0)}_{\rm eff}&=i\alpha\big( f  + h'\,   S +\hat f'\, Y\big)
} 
for the effective K\"ahler- and superpotential,
with the model dependent order one parameters $A,\alpha,\kappa\in \mathbb
R$. Just for simplicity, we were choosing $\alpha=\beta=\gamma$.
Including just these single  first order terms in $W_{\rm eff}$ implicitly
means that we have assumed that possible higher order polynomial terms
like $Y^n$ are either not present or are subleading.
Whether the concrete form of realistic periods admit
such a choice remains to be seen.

For the model \eqref{effaxionmodelb} we find a Minkowski minimum at 
\eq{ 
         \Sigma&=\hat f' \, \zeta_0+h'\, c_0 =0 \\[0.2cm]
        s&=s_0={1\over h'}\sqrt{ f^2 -A\, \hat f'^2+ \kappa  f\hat f'}\\[0.1cm]
        {\rm Re}(Y)&=y_0= {1\over \hat f'}\bigg( \!-f+\sqrt{ f^2 -A\, 
             \hat f'^2+ \kappa  f\hat f'}\,\bigg)\,.
}
where $Y=y+i\zeta$.
Note that for $f/\hat f'\gg 1$ these expressions simplify drastically
\eq{ 
         \Sigma =0\,,\quad\quad
        s_0={f\over h'}\,,\quad\quad
        y_0={\kappa\over 2}+O\!\left({\textstyle {\hat f'\over f}}\right)\,.
}
Therefore,  for $ f/h'\gg 1$ and $\kappa<1$ the values of
the moduli lie in the perturbative regime of  small string coupling and
small complex structure modulus $Y$. In the following, 
we analyze this model in more detail with special focus on the
axion sector. For simplicity we choose $\kappa=0$ so that
\eq{ 
\label{min_align}
      \Sigma =0\,,\quad\quad
        s_0={f\over h'}\,,\quad\quad
        y_0=-{A \hat f'\over 2f}+O\!\Big({\textstyle ({\hat f'/ f})^2}\Big)\,.
}
Note that the second axion $\Theta$ is still massless at this level.
It is clear that once additional terms like e.g. $\Delta W=iB\, Y^2$ are
present in the superpotential, $\Theta$ also gets stabilized with a mass that is governed by
the  parameter $B$. For $B$ of order one, all four fields will have
the same order of masses, but for a parametrically smaller value of 
$B$ the axion $\Theta$ will be the lightest state of all complex
structure and axio-dilaton moduli.  

Recall that, after integrating out $Z$, one also 
gets an exponential  term like $\exp(-{\textstyle {2\pi \over f}} h
S)$ in $W$. As its size depends on a different modulus, this term can in principle compete with the higher
order polynomial terms $Y^n$. For illustrative purposes, let us consider in
the following section the possible moduli stabilization scheme,
once we include only this exponential term in $W$ and
assume that polynomial terms are either absent or subleading.
We understand that this is a very strong assumption that needs
to be tested for concrete CY manifolds (e.g.  along the line
reported in \cite{Bizet:2016paj}).

\subsection{Axion alignment}

As we have seen, only one linear combination of the two axions is
stabilized by terms appearing linearly  in $W$.
We now  analyse the phenomenological effective supergravity model \eqref{effaxionmodelb}
\eq{
\label{finalchallenge}
         W_{\rm eff}&=W^{(0)}_{\rm eff}+W^{(1)}_{\rm eff}\\
&=i\alpha\big( f+  h'\,    S + \hat f'\,  Y\big)
         +{f \hat C\over 2\pi i} \, \exp\!\Big(\! -{\textstyle {2\pi \over f}} \big(h S
              + \hat f Y\big)\Big)\,.
} 
Here we have included a term
$\exp(- {\textstyle {2\pi \over f}} \hat f Y)$ which is induced from a coupling $\hat f Z Y$
in the original superpotential.

Let us now derive an effective scalar potential $V_{\rm eff}$ for the so far
unstabilized axion appearing in the exponent of
\eqref{finalchallenge}. If we naively integrate out the already
stabilized moduli via
\eq{
        D_S W_{\rm eff}|_{S_0,Y_0}&=c_S\, |Z| \exp\!\Big(\! -{\textstyle {2\pi i\over f}} \Theta\Big)\,,\\
    D_Y W_{\rm eff}|_{S_0,Y_0}&=c_Y\, |Z| \exp\!\Big(\! -{\textstyle {2\pi i\over f}} \Theta\Big)
}
with 
$\Theta=\big(h c  + \hat f \zeta\big)$
 and $c_S,c_Y\ne
0$, we realize that $V_{\rm eff}|_{S_0,Y_0}$ does not depend on the
axion $\Theta$ and is non-vanishing at order
$O(|Z|^2)$. However, this is not what one expects. In the true vacuum (at
order $O(|Z|^2)$) the vacuum energy should be  vanishing and
the remaining axion $\Theta$ should be stabilized at $\Theta=0$.

Indeed to see this, we first have to take the backreaction of the exponential
term on the stabilization of the saxions into account.
Perturbing around the leading order values $y_0,s_0$ by   $\Delta y_0\sim O(|Z|) $ and  $\Delta s_0\sim
O(|Z|) $, we  fix $\Delta y_0$ and $\Delta s_0$  by requiring 
$D_S W_{\rm eff}|_{S_0,Y_0,\Theta=0}=D_Y W_{\rm
  eff}|_{S_0,Y_0,\Theta=0}=0+O(|Z|^2)$.
For $\kappa=0$, $ f/\hat f'\gg 1$ and at leading non-vanishing order in $|Z|$
we find
\eq{
	\Delta s_0 \sim - \frac{f}{2 \pi \alpha h'} \Big(1 + \frac{4 \pi h}{h'}\Big) \, |Z| \, , \qquad
	\Delta y_0 \sim - \frac{A \hat f}{2 \alpha f} \, |Z| \,  \, .
	}
This backreaction induced shift of the vacuum leads to the correct effective scalar potential
\eq{
     V_1&=G^{S\ov S}\, D_S W\, D_{\ov S} \ov W={|Z|^2\,\over
      2 \pi^2}\ f^2 \Big(1+{4\pi h\over h'}\Big)^2\,\,
    \Big(1 - \cos\big({\textstyle {2\pi\over f}} \Theta\big)\Big)\\[0.2cm]
      V_2&=G^{Y\ov Y}\, D_YW\, D_{\ov Y} \ov W={|Z|^2\, A\over
       4 \pi^2}\, \hat f'^2\,\Big(1+\frac{4\,\pi \hat f}{\hat f'}\Big)^2\,
    \Big(1 - \cos\big({\textstyle {2\pi\over f}} \Theta\big)\Big) \, ,
}
so that in the regime  $f/\hat f'\gg 1$ and $h/h'\gg 1$ we eventually
obtain the effective potential for $\Theta$
\eq{
\label{nicecospot}
     V_{\rm eff}&=e^{K_{\rm eff}}\,(V_1+V_2) \sim {4 |Z|^2\,\over
      A {\cal V}^2}\ {f h^2 \over h'}\,\
    \Big(1 - \cos\big({\textstyle {2\pi\over f}} \Theta\big)\Big)
}
Note that the potential exhibits the expected Minkowski minimum at $\Theta=0$.

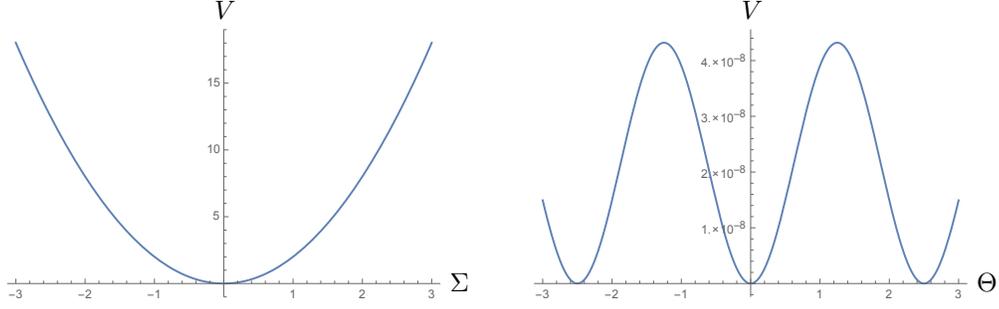
\begin{figure}[!ht]
 \centering
\includegraphics[width=0.4\textwidth]{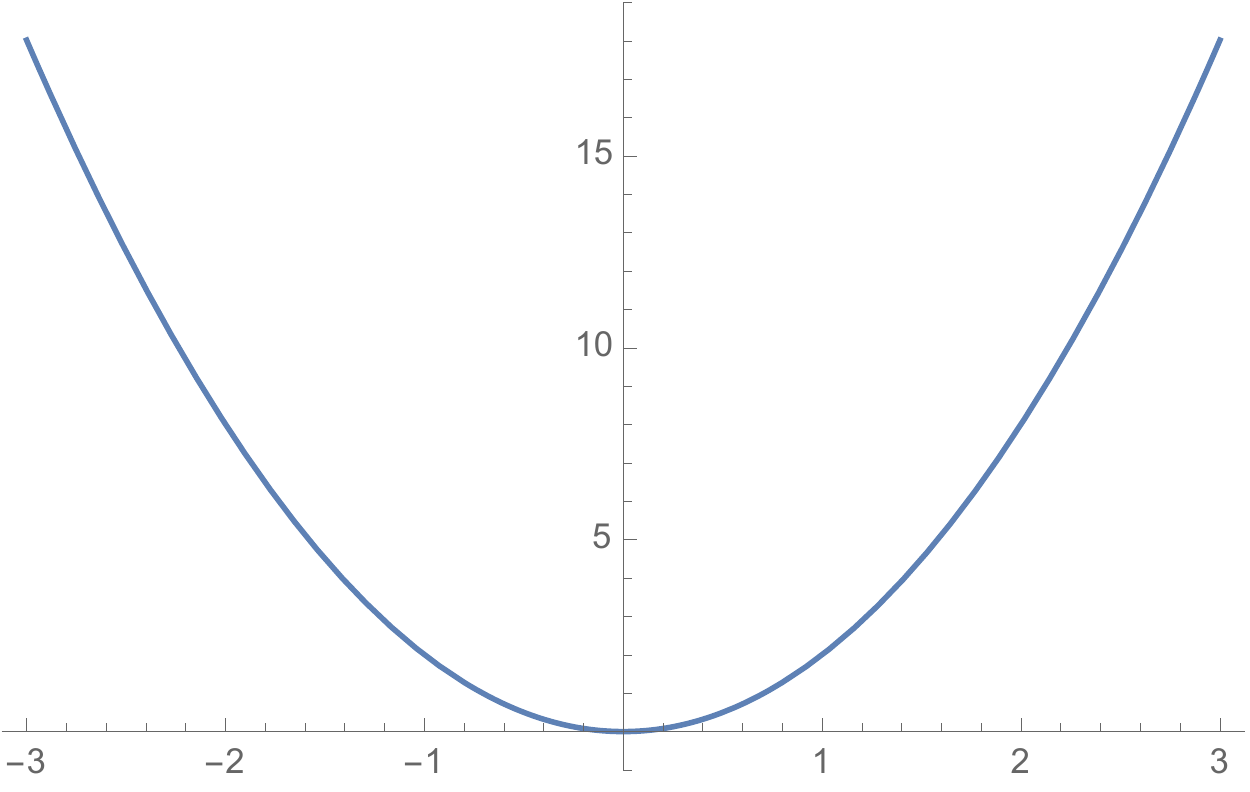}
 \begin{picture}(0,0)
      \put(-90,108){\footnotesize $V$}
    \put(-,5){\footnotesize $\Sigma$}
  \end{picture}
\hspace{0.8cm}
\includegraphics[width=0.4\textwidth]{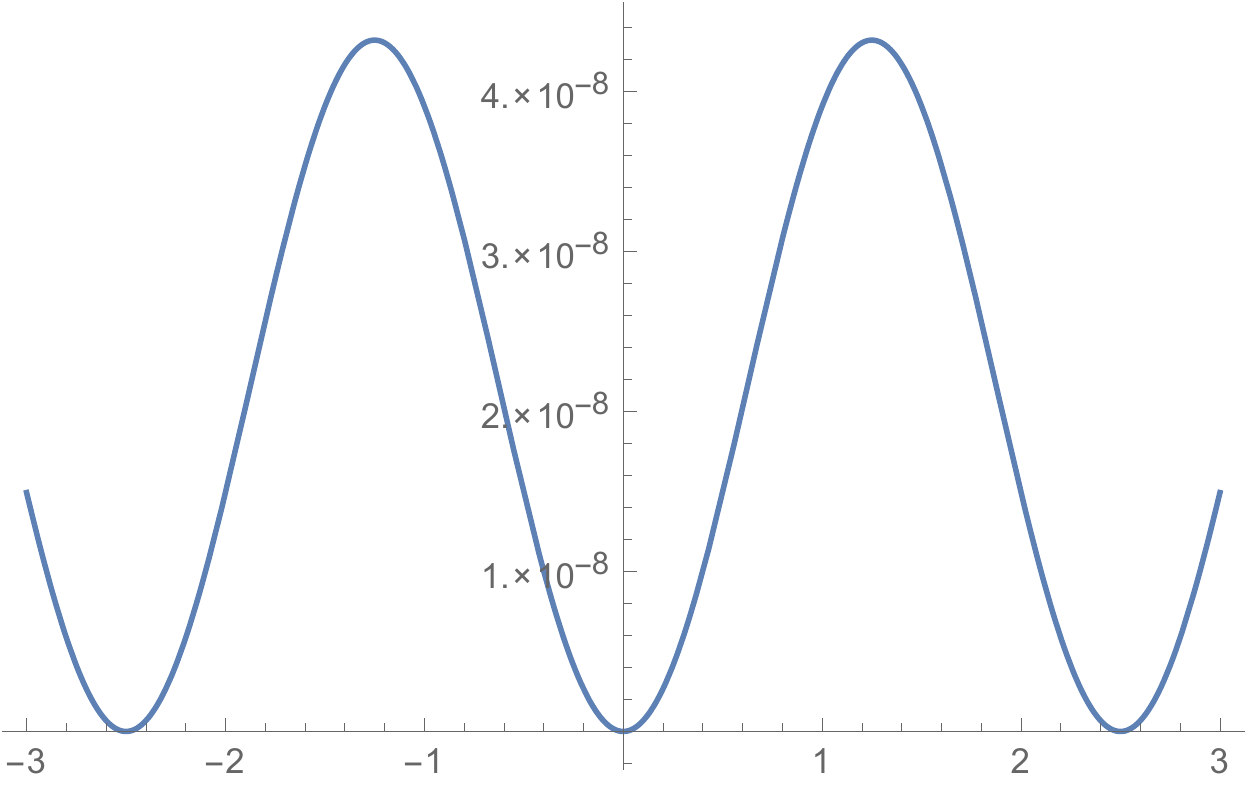} 
\begin{picture}(0,0)
   \put(-90,108){\footnotesize $V$}
   \put(-1,5){\footnotesize $\Theta$}
  \end{picture}
\caption{\label{fig_A} Scalar potential for the two axions $\Sigma$
  and $\Theta$ for $f=10$, $h'=\hat f'=1$, $h=-\hat f=2$, $\hat C=1$
  and $A=0.1$.}
\end{figure}

As a check of our approach,
in figure \ref{fig_A} we have plotted the axion dependence of
the full scalar potential for the saxions fixed at the Minkowski
minimum. It nicely shows the periodic form of the potential
\eqref{nicecospot} and that the height of the potential in 
$\Theta$ direction is hierarchically smaller than in $\Sigma$
direction.

\vspace{0.5cm}

\noindent
In order to apply the model at hand to axion inflation, it remains to rewrite the potential in terms of the canonically normalized field $\tilde \Theta$ which we then employ as inflaton.
Due to the diagonal structure of the K\"ahler metric, one finds in the regime $f/f' \gg 1$
\eq{
	 \tilde\Theta={h'\over \sqrt{A} \hat f'} \Theta \, ,
	 }
such that the canonically normalized inflaton potential is given by
\eq{
\label{pot_inflaton}
	V_{\rm eff} \, = \, {4 |Z|^2\,\over
      A {\cal V}^2}\ {f h^2 \over h'}\,
    \Big(1 - \cos\big({\textstyle {2\pi \sqrt{A} (h \hat f' - h' \hat f) \over f h'}}\tilde \Theta\big)\Big) \,
     \equiv \, V_0 \Big( 1 - \cos \big( {\textstyle \frac{\tilde \Theta}{f_{\tilde \Theta}}} \big) \Big) \, .
	}
The axion decay constant $f_{\tilde \Theta}$ signalizes the appearance of an alignment mechanism as
\eq{
	f_{\tilde \Theta} = \frac{f}{2 \pi \sqrt{A}} \, \frac{h'}{h \hat f' - h' \hat f} \, .
	}
By aligning the fluxes as $(h \hat f' - h' \hat f)<h'$
we can obtain an axion decay constant larger than one.
This is very reminiscent of the KNP-axion alignment mechanism \cite{Kim:2004rp}, the main difference being
that one linear combination of axions is fixed by fluxes at linear
order in the fields and only the second combination by instanton-like
terms.

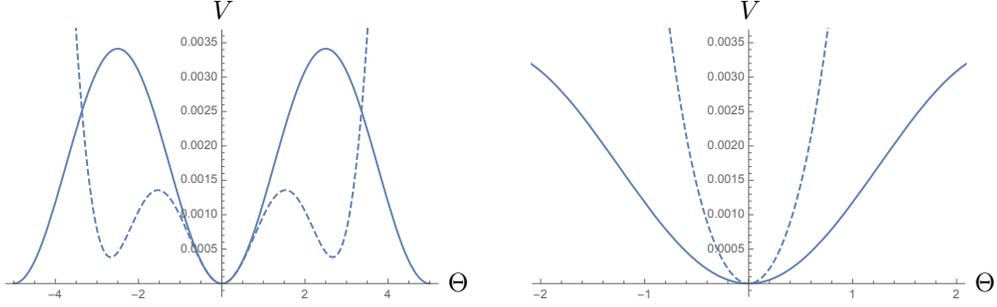
\begin{figure}[!ht]
 \centering
\includegraphics[width=0.4\textwidth]{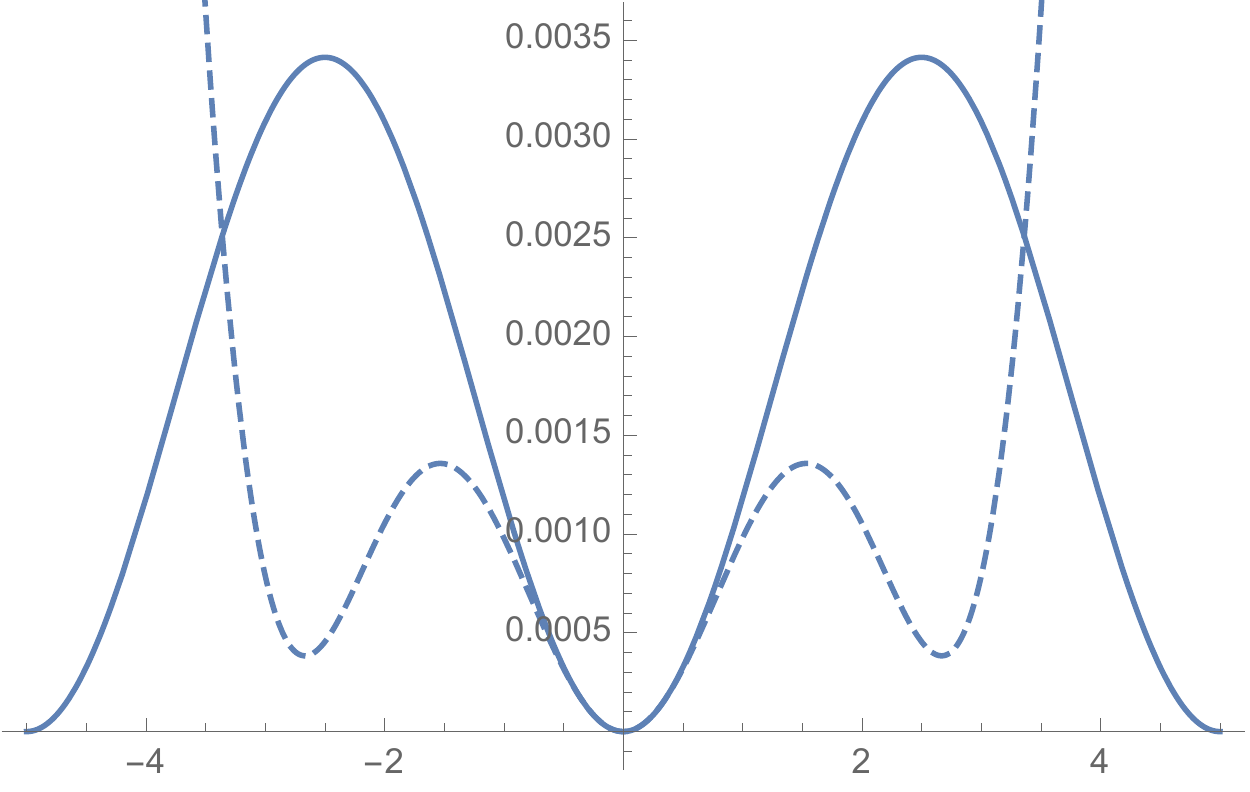}
 \begin{picture}(0,0)
      \put(-90,108){\footnotesize $V$}
    \put(-,5){\footnotesize $\Theta$}
  \end{picture}
\hspace{0.8cm}
\includegraphics[width=0.4\textwidth]{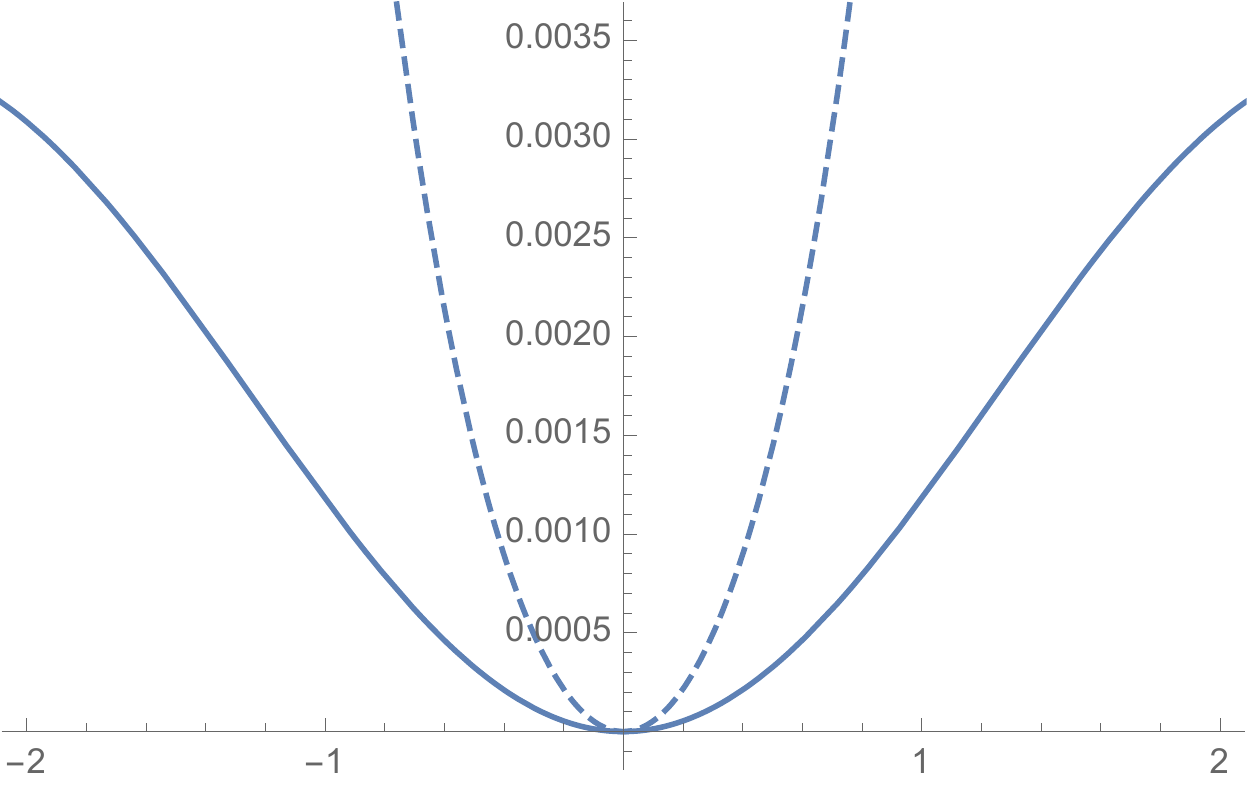} 
\begin{picture}(0,0)
   \put(-90,108){\footnotesize $V$}
   \put(-1,5){\footnotesize $\Theta$}
  \end{picture}
\caption{\label{fig_D} Scalar potential (dashed lines) for the  axion 
  $\Theta$ for $f=10$, $h'=\hat f'=1$, $h=-\hat f=1$, $\hat C=1$,
  $A=0.1$
   and $B=0.01$ in the left-handed  plot and $B=0.1$ in the
   right-handed  plot. For comparison, the solid lines show the potential for $B=0$.
}
\end{figure}

Coming back to our assumptions about the suppression of polynomial
terms in $W$, we consider the superpotential
\eq{
\label{finalchallengeb}
         W_{\rm eff}=i\alpha\big( f+  h'\,    S + \hat f'\,  Y\big) +iB Y^2
         +{f \hat C\over 2\pi i} \, \exp\!\Big(\! -{\textstyle {2\pi \over f}} \big(h S
              + \hat f Y\big)\Big)\,
} 
and in figure \ref{fig_D} display the new effective potential for
$\Theta$ (by  dashed lines) for two different values of the parameter
$B$.

\vspace{0.5cm}

The left-handed plot shows that here $B$ is sufficiently small so
that in the direct neighborhood of $\Theta=0$ the potential
is dominated by the exponential term.  Whereas, in the  right-handed
plot $B$ is so large that the mass of $\Theta$ comes from the quadratic
term. As said, we are not claiming that higher order terms are really
suppressed for a concrete CY, but just want to show which kind
of scenarios are in principle possible for moduli stabilization close
to the conifold. We would not be surprised if once again
a concrete string theoretic proposal for realizing large field inflation
fails as one loses control over certain dangerous terms.
In this spirit we proceed with discussing the purely exponential case.

\subsubsection*{A comment on the WGC}

Even though the effective exponential terms in the superpotential do 
not directly arise from instanton contributions, one can ask whether
they satisfy a generalized version of  the weak gravity conjecture.
For instantons the conjecture says that the product of the instanton
action times the axion decay constant has to be smaller than one
\eq{
                     S_{\rm inst}\, f_{\rm inst} \le 1\,.
}
If this is to be satisfied for the instanton with the lowest action, one has
the strong version of the WGC. Apparently, this is violated for the
aligned axion $\tilde \Theta$ whose action can be written as
\eq{
S_{\rm inst}= {2\pi\over f}(h s_0+\hat f y_0)\sim {2\pi\over f} h
s_0\sim {2\pi h\over h'}\,,
}
so that in the case of  alignment one gets
\eq{
            S_{\rm inst}\, f_{\tilde\Theta}\sim 
            \frac{fh}{\sqrt{A} (h \hat f' - h' \hat
              f)} >1\,.
}
The \emph{mild} form of the WGC says  that for an axion with decay
constant $f_{\tilde\Theta}$ there is \emph{some} instanton with
instanton action $S_{\rm inst}^{(2)}$
satisfying  the WGC condition. 
Therefore, to satisfy the mild WGC, it is sufficient to have a
situation where the
contribution of these two instantons reads as 
\eq{
	V \sim  e^{-2\,S_{\rm inst}} \, \Big( 1 - \cos \big(
            {\textstyle \frac{\tilde \Theta}{f_{\tilde \Theta}}} \big) \Big) +
	e^{-2\,S_{\rm inst}^{(2)}} \, \Big( 1 - \cos \big(
            {\textstyle \frac{k \, \tilde \Theta}{f_{\tilde \Theta}}} \big) \Big) \, ,
	}
with $k\in \mathbb{Z}$. For sufficiently large $k$, the axion decay
constant $f_{\tilde \Theta}/k$ can be sub-Planckian and for $S_{\rm
  inst}<S_{\rm inst}^{(2)}$ the first term can still be dominant and realize
inflation. This loop-hole was pointed out in \cite{Rudelius:2015xta,Brown:2015iha} and has been
realized for complex structure aligned inflation in \cite{Hebecker:2015rya}.

In our case the situation is similar, where the second exponential
contribution could arise from  a single  $D(-1)$-instanton.
Its action and decay constant
is
\eq{
             S_{D(-1)}={2\pi s_0}={2\pi f \over h'}\,,\qquad
          f_{D(-1)}={h'\over 2\pi \sqrt{A} \hat f'}={f_{\tilde\Theta}\over k}
}
with $k=(f\, \hat f')/(h \hat f' - h' \hat f)$. With the denominator being equal
to one in the case of alignment, $k$  is a large integer. Moreover,
for $f/h>1$ the $D(-1)$ instanton action is subleading and inflation
can still occur.

\subsubsection*{Mass scales}

Let us finally compute the different mass scales to confirm our
various effective approaches and in particular justify integrating out
massive fields at several stages of our computation.
Let us denote by $M_{\rm mod}$ the masses of the moduli $\Sigma$, $s$ and $\text{Re} (Y)$ in the minimum \eqref{min_align}.
Up to numerical prefactors of order $O (1)$, the mass eigenvalues of the canonically normalized mass matrix $(M^2)^i_{\ j} = \frac{1}{2} G^{i k} \partial_k \partial_j V$ are given by 
\eq{
	M^2_{\rm mod} = \frac{f^2 g_s}{ A \mathcal V^2} \, .
	}
In addition one can then read off the mass of the canonically normalized inflaton $\tilde \Theta$ from $V_{\rm eff}$ in equation \eqref{pot_inflaton}
\eq{
	M^2_{\tilde \Theta} \, = \, \frac{V_0}{f_{\tilde \Theta}^2} \,
        M^2_{{\rm Pl}}\sim \frac{|Z|^2}{f\,\mathcal V^2} \, M^2_{\rm Pl}
                                \, .
	}
Hence the mass of the inflaton is exponentially suppressed relative to the mass
of the other moduli stabilized at tree-level. Recall that the latter are
also exponentially suppressed relative to the conic complex structure modulus $Z$, such that we obtain a mass hierarchy of the form
\eq{
	M_{\tilde \Theta} \, < \, M_{\rm mod} \, < \, M_{Z} \, .
}
The mass scale of inflation can be read off from \eqref{pot_inflaton} as
\eq{
               M_{\rm inf}^2\sim V_0^{1\over 2}\sim {f^{1\over 2}
                 |Z|\over {\cal V}}\,.
}
Therefore, with $g_s\sim 1/f$ we obtain
\eq{
          {M_{\rm inf}^2\over M^2_{\rm mod}}\sim { ({\cal V} |Z|^2)\over
            f^{1\over2} |Z|}\,\qquad {\rm and}\qquad
       {M_{\rm inf}^2\over M^2_{Z}}\sim ({\cal V} |Z|^2)
       {|Z|\over f^{1\over 2}}
}
so that in the regime ${\cal V} |Z|^2\gg 1$ the inflationary scale is larger
than the moduli masses, but for sufficiently small $|Z|$ can
be lower than the mass of the conic complex structure modulus.
Therefore,  for correctly describing the dynamics in the slow-rolling
phase one can use the effective four-dimensional SUGRA theory after integrating out
the conic modulus $Z$. The backreaction of the inflaton on the
remaining moduli  is expected to lead to a (welcomed) flattening of
the quadratic inflaton potential \cite{Dong:2010in}.
Note that in this respect this model behaves better than 
the ones constructed in the framework of the flux scaling scenario
\cite{Blumenhagen:2015kja},
where generically the inflationary mass scale was even larger than the
KK scale.

So far we did not stabilize the K\"ahler moduli for this inflationary
model. Let us now assume that we can employ the large volume scenario
and estimate the appearing mass scales as in section \ref{sec_conLVS}.
We ignore possible subtleties about the order of integrating out for the moment.
Since the non-supersymmetric LVS minimum is of AdS type, we also have to assume a proper
uplift mechanism. As a first rough estimate,  in table
\ref{table_B} we list   all relevant mass scales.

\vspace{0.2cm}

\begin{table}[ht]
\centering
\renewcommand{\arraystretch}{2.0}
\begin{tabular}{|c|c|}
  \hline
   Scale & $(\text{Mass})^2$ in $M^2_{\rm Pl}$\\
 \hline\hline
  string scale $M^2_{\rm s}$ & $\displaystyle
  \frac{1}{f^{1/2}\, {\cal V}}$ \\
  Kaluza-Klein scale $M^2_{\rm KK}$ & $\displaystyle \frac{1}{{\cal V}^{4/3}}$ \\
  conic c.s. modulus $M^2_Z$ &
  $\displaystyle 
     {f \over {\cal V}^2 |Z|^2 }$ \\
  inflationary mass scale  $M^{2}_{\rm inf}$ & $\displaystyle 
   \frac{f^{1/2} |Z|}{\mathcal V}$\\
    other moduli
  $M^2_{\rm mod}$ & $\displaystyle  \frac{f}{\mathcal V^2}$ \\
   gravitino mass
  $M^2_{\rm 3/2}$ & $\displaystyle  \frac{f}{\mathcal V^2}$ \\
 large K\"ahler modulus $M^2_{\tau_b}$ &
  $\displaystyle \frac{ f^{5/2}}{ \mathcal{V}^3 }$  \\
  inflaton $M^2_{\tilde \Theta}$ & $\displaystyle
   \frac{|Z|^2}{f \, \mathcal V^2}$\\[4mm]
\hline
     \end{tabular} 
    \caption{\label{table_B}  Moduli masses and scales with $g_s\sim 1/f$.}
\end{table}

\noindent
From the table we extract the relation
\eq{
       {M^2_{\tilde \Theta}\over  M^2_{\tau_b}}\sim {{\cal V} |Z|^2\over
         f^{7/2}}
}
so that in the reliable SUGRA regime with ${\cal V} |Z|^2\gg 1$ we
generically find that the large K\"ahler modulus is lighter than the potential
inflaton. Of course this spoils single field inflation and reflects a
problem that seems to be very generic for complex structure moduli
inflation \cite{Blumenhagen:2014nba,Hebecker:2014kva}.
One can derive the relation
\eq{
       {M^2_{\tilde \Theta}\over  M^2_{\tau_b}}\sim 
      {M^2_{\rm s}\over  f^2 M^2_Z}
}
so that  in principle for $M_{\rm s}/ M_Z\sim 5-8$ one can get 
that the axion $\tilde\Theta$ is the lightest mode for $f\sim 10$.
Of course here one is at the boundary of control and numerical factors
matter.
Thus, the generic  hierarchy of scales is of the form
\eq{
	 M_{\tau_b}< M_{\tilde \Theta} \, < \, M_{\rm mod} \, < \, M_{\rm inf} \, \sim \, M_{Z} \,
	< \, M_{\rm KK} \, < \, M_{\rm s} \, < \, M_{\rm Pl} \, ,
	}
guaranteeing  parametric control over the mass scales in our effective SUGRA description. 
It is not excluded that by a certain choice of the flux
$f$ one can get that the axion is in principle the lightest mode. In this case
a more detailed analysis is necessary, as one cannot first stabilize
the axion $\tilde \Theta$ and then integrate out the K\"ahler moduli.

Summarizing, in this corner of the string theory
landscape we managed to design a string motivated effective
supergravity model that features an alignment
mechanism providing an axion that has an effective decay constant
larger than one. The axion sector still satisfies the mild
form of the WGC.
However, this model will probably  fail at the end,
as we had to make  very strong assumptions about higher order polynomial corrections to $W$ and
because it is not yet clear whether one can get the K\"ahler heavier
than the inflaton.

\section{Conclusions}

In this paper we studied moduli stabilization close to the 
conifold locus in the complex structure moduli space.
This is an attractive corner of the string theory landscape,
as the K\"ahler potential features axionic shift symmetries
that are interesting for string theory realizations of large field
inflation.

Following the early ideas of GKP, after computing the periods
close the conifold locus, by turning on appropriate fluxes, we managed to dynamically
stabilize the conic complex structure modulus at $|Z|\ll 1$ and the
other complex structure moduli and the axio-dilaton in their
perturbative regime. As a self-consistency condition for the
use of the GVW effective field theory we found that the total
volume must be exponentially large  precisely  guaranteeing
the absence of significant warping. We showed
that a combination with the large volume scenario is 
possible, providing a consistent hierarchy of mass scales.

Moreover, we found that even in the absence of warping certain
modes still show exponential mass hierarchies, that can be traced
back to the appearance of log-terms in the periods. Let us emphasize
that these are not the modes whose mass gets red-shifted in the
region of strong warping. After integrating out
the heavy conic complex structure modulus, one generates
exponential, instanton-like terms in the superpotential.
As demonstrated, these can be further exploited for moduli
stabilization.

Being equipped with new mechanisms to generate exponential hierarchies
we approached the important problem of designing a 
string theory motivated model of axion inflation featuring
consistent moduli stabilization. Ignoring first the K\"ahler moduli and
following the main idea of \cite{Hebecker:2015rya} we managed to construct a model including an aligned axion
with axion decay constant larger than one. However, in the
construction
quite  strong assumptions about the size of higher order corrections
were made and 
we were not able
to also freeze the K\"ahler moduli via the LVS such that we really
end up with a fully controllable model of single field inflation.
It
would be interesting to see whether, as reported in \cite{Bizet:2016paj}, the higher order corrections
to the periods do  spoil some of our findings. 


It is not clear to us  whether those effective exponential terms 
are related to true instanton contributions on the mirror side and
whether a form of the weak-gravity conjecture applies to them.
Recall that they  only became visible after integrating out the conic
modulus $Z$. Thus, they might indeed provide a loop-hole in the
axionic version of the WGC.

Finally, let us just state that the
landscape is  rich and still might contain new mechanisms for hierarchical
moduli stabilization that are waiting to be exploited for various
applications in string phenomenology and string cosmology.

\vspace{0.8cm}
\noindent
\emph{Acknowledgments:} We are grateful to Shanta de Alwis, Anamaria
Font, Thomas Grimm, Kepa Sousa and Irene Valenzuela  for
discussions and to Michael Fuchs for collaboration during early
stages of this work.
Moreover, we thank Nana Cabo Bizet, Oscar Loaiza-Brito and Ivonne Zavala for pointing
out a numerical mistake in an earlier version of this paper.


\clearpage
\bibliography{references}  
\bibliographystyle{utphys}


\end{document}